\documentstyle[aps]{revtex}
\usepackage{amsfonts}
\usepackage{a4wide}
\usepackage{epsfig}
\newfont{\cyr}{wncyr10}
\newcommand{\news}{\setcounter{equation}{0}}
\newcommand{\be}{\begin{equation}}
\newcommand{\ee}{\end{equation}}
\newcommand{\bea}{\begin{eqnarray}}
\newcommand{\eea}{\end{eqnarray}}
\newcommand{\bean}{\begin{eqnarray*}}
\newcommand{\eean}{\end{eqnarray*}}
\newcommand{\V}{{\cal V}(\chi\chi^*)} 
\font\upright=cmu10 scaled\magstep1
\font\sans=cmss12
\newcommand{\lsim}{\mathrel{\raise.3ex\hbox{$<$\kern-.75em\lower1ex\hbox
{$\sim$}}}}
\newcommand{\ssf}{\sans}
\newcommand{\stroke}{\vrule height8pt width0.4pt depth-0.1pt}
\newcommand{\Z}{\hbox{\upright\rlap{\ssf Z}\kern 2.7pt {\ssf Z}}}

\newcommand{\C}{{\rlap{\rlap{C}\kern 3.8pt\stroke}\phantom{C}}}
\newcommand{\R}{\hbox{\upright\rlap{I}\kern 1.7pt R}}
\newcommand{\CP}{\C{\upright\rlap{I}\kern 1.5pt P}}
\newcommand{\PP}{\hbox{\upright\rlap{I}\kern 1.5pt P}}

\newskip\humongous \humongous=0pt plus 1000pt minus 1000pt

\newif\ifdtup

\relax

\begin{document}
\title  {Topological symmetry breaking and confinement of anyons}
\bigskip

\author{P.W. Irwin 
and
M.B. Paranjape
}
\address{Groupe de Physique des Particules, D\'{e}partement de Physique,
Universit\'{e} de Montr\'{e}al, \\
C.P. 6128 succ. centre-ville, Montr\'{e}al, Qu\'{e}bec, H3C 3J7, Canada \\
{\scriptsize irwin@lps.umontreal.ca, paranj@lps.umontreal.ca } 
}

\date{May 1999}
\maketitle

\begin{abstract}
We study the behaviour of Wilson and 't Hooft loop operators for the 2+1 dim.
Abelian-Higgs model with Chern-Simons term. The phase of topological symmetry 
breaking
where the vortex field condenses, found by Samuel for the model in the absence
of the Chern-Simons term, persists in its presence.  In this phase, the 
topological linking of instantons, which are configurations of closed vortex 
loops, with the Wilson loop on one hand and with the 't Hooft loop on the 
other hand gives rise to a long-range, logarithmic, confining potential 
between electric charges and magnetic flux tubes.  This is surprising since
the gauge field is short range due to the Chern-Simons term.  Gauss'
law forces the concomitance of charge and magnetic flux, hence the 
confinement is actually of anyons. 
\end{abstract}

\bigskip

\section{Introduction}
\news
\ \indent 
The Abelian-Higgs model 
has long been a model of wide interest from condensed matter to
particle physics.  It is the effective theory for the description of
superconductivity 
with a charged condensate. It also provides the simplest example 
of the Higgs' mechanism where spontaneous breaking of a gauged
symmetry evades the Goldstone theorem and provides for a purely massive
spectrum. The theory is expected to have 
two phases, the Higgs (or superconducting) symmetry broken phase and
the Coulomb phase. The expectation value of the Wilson loop operator which
measures the potential between external electrically charged
particles, distinguishes between the two phases.  In lower dimensions
it is possible that 
non-perturbative phenomena disturb this straightforward conclusion.  
In 1+1 dimensions instanton effects actually conspire to restore 
the symmetry in the putative Higgs phase, and indeed there reappears
only a Coulomb 
phase with however a distinctly different strength for the 
linear confining potential \cite{CDG}. The result
is essentially topological, the Wilson loop divides space-time into
two disjoint parts, the Wilson loop integral measures whether the
instanton is (topologically) inside the loop or outside the loop.  In
2+1 dimensions \cite{Sam}, nonperturbative 
instanton contributions to the expectation value of the Wilson loop
were computed.  In the Higgs region topological solitons exist in the 
spatial plane 
as vortices, their continuation to three dimensional Euclidean space are 
vortex loops which serve as the instantons. Their contribution to the 
path integral 
requires summing over all configurations of vortex loops.  Again the
contribution is essentially topological, the Wilson loop integral
measures the topological linking number of the vortex loop with 
the Wilson loop.  

On a lattice of fixed spacing one finds the 
contribution of loops of length $L$ to be approximately $e^{-mL}\mu^L$,
where $m$ is the vortex mass per unit length, and $\mu$ is an intrinsic 
parameter such that $\mu^L$ counts the degeneracy of loops of length $L$.
A phase transition occurs when $m-ln\,\mu\sim 0$ 
where vortex loops of all lengths become important, 
which has been seen to occur in lattice simulations 
of the model, see e.g. \cite{Loop}. 
A partition function of a gas of loops can be converted to a 
effective field theory of a scalar field called the vortex field.
Condensation of the vortex field corresponds to the ``spaghetti vacuum''
and is called topological symmetry breakdown. 
The vacuum is populated of vortex/anti-vortex pairs 
restoring the $U(1)$ symmetry giving back the Coulomb 
phase.  In the region where the vortex field condenses
the vortex loops induce a nonperturbative logarithmic confining 
force between external charged particles \cite{Sam}. 
If the external particles have charge $q$, the strength 
of the logarithmic potential is periodic in $q$ with period $e$, 
the charge of the elementary scalar particles in the theory, since
they can screen the external 
particles when $q$ is a multiple of $e$. 
  
In 2+1 dimensions or more, one can also look at the 
behaviour of the 't Hooft loop 
operator \cite{tHooft}. It is defined by the action of
a singular gauge transform on the fields, and
at each time slice it
creates a vortex/anti-vortex pair. 
The behaviour of the 't Hooft loop expectation value describes 
the inter-vortex potential.
The perturbative and vortex loop configurations give a perimeter 
law dependence to the 't Hooft loop for both phases of the theory.

If the Chern-Simons term (CS), with coefficient $\kappa$ 
is included we ask what is its effect on the Wilson and 
't Hooft loop expectation values? The immediate problem that arises 
is that the CS term becomes imaginary in Euclidean space. 
Since the action is complex, the Euler-Lagrange 
equations over determine the system and there is in 
general no solution to the field equations (in terms of real fields). 
There do exist complex solutions however,
for small $\kappa$ we expect 
that the real vortex loop configurations found at $\kappa=0$ 
will continue to be the most relevant configurations. 
Our approach is thus the same as \cite{AHPS}, who considered the effect 
of the CS term to the monopole contribution in the Georgi-Glashow model. 
In fact we can show complex vortex solutions exist in the present case but 
we do not use them here.

We find that for the 2+1 dimensional Abelian-Higgs model the addition 
of the CS term does not significantly alter the vortex loop
contribution to the Wilson loop expectation value for small
$\kappa$. The behaviour of 
the 't Hooft loop is however significantly 
altered. In a manner similar to the case 
without the CS term, there is a phase transition at the point where the 
entropy factor of the vortex loops overcomes the Boltzmann factor. 
When the vortex field condenses, vortex loops
give a logarithmic potential between both the 
external electric charges and vortices.
This result is very surprising given that the photon is now massive 
and perturbatively no long range forces exist.  
It is unclear to us whether this behaviour persists throughout
the unbroken symmetry phase. For the vortex
entropy factor to overcome the Boltzmann factor  
the classical mass of the vortex must be smaller than some critical 
value where the phase transition occurs. But topological vortex 
configurations exist classically only for a certain range of parameters. 
The mass of the vortex can be continuously reduced to zero, in this limit 
the topological stability of the vortex is lost. This corresponds to the 
region where the gauge symmetry is unbroken classically. Since we cannot 
construct topological vortex configurations here, our methods do not 
extend to this region. We speculate that the confining logarithmic potential 
continues throughout the symmetry unbroken phase.

In Section 2 we describe the model and its vortex solutions.
We then describe the contributions of the vortex loops 
to the path integral and discuss how to treat the path 
integral with the inclusion of the complex CS term in Euclidean 
space. In Section 3 we compute the instanton contribution
to the Wilson and 't Hooft loop expectation values and  
conclude with comments.

\section{The Abelian-Higgs model with CS term}
\news
\ \indent 
The Abelian-Higgs model with CS term in (2+1) dimensions has 
Lagrangian density given by
\be 
{\cal L}=\left\{-\frac{1}{4}F_{\mu\nu}F^{\mu\nu} +\frac{1}{2}
D_{\mu}\phi\,(D^{\mu}\phi)^*-
\frac{\lambda}{4}(|\phi|^2-\eta^2)^2+\frac{\kappa}{4}\epsilon
^{\mu\nu\rho}A_{\mu}F_{\nu\rho} 
\right\}\;,\label{Ldens}
\ee
here $\phi$ is a complex scalar field, 
$A_{\mu}$ is a $U(1)$ gauge potential,
$D_{\mu}=\partial_{\mu}-ieA_{\mu}$,
$F_{\mu\,\nu}=\partial_{\mu}A_{\nu}-\partial_{\nu}A_{\mu}$ 
($\mu$, $\nu=0,\,1,\,2$)
with $\lambda$, $e$, $\kappa$ and $\eta^2$ coupling constants.
$\kappa$, $\lambda$, $\eta^2$ and $e^2$
each have the dimension of mass. We are considering noncompact
$U(1)$ gauge theory, there is no quantization condition on $\kappa$.
When $\kappa=0$, if $\eta^2\leq 0$ the gauge symmetry is unbroken and 
the perturbative spectrum consists of a 
massless photon and a massive charged 
scalar particle of charge $e$. If $\eta^2>0$ the gauge symmetry is 
spontaneously broken and the particle content is given 
by a massive vector particle and a massive neutral scalar particle. 
In addition classical time independent solitons exist in the form of 
Nielson-Olesen vortices. The 
vortices carry a topologically conserved integer
valued charge usually called the vortex number. 
The vortex mass $m$ is approximately equal to  
$\eta^2$. For $\eta^2\leq 0$ the  
topological vortex solution no longer exists. 

If the CS term is now included, $\kappa\neq 0$, the theory 
is altered in that the photon is massive even 
when $\eta^2\leq 0$. The CS term automatically gives the photon a mass 
proportional to $\kappa$, this is called topologically massive QED.  
In the unbroken phase when $\eta^2\leq 0$ the perturbative 
particle spectrum consists of a massive 
charged scalar field and a massive gauge particle. 
When the $U(1)$ symmetry breaking 
occurs, for $\eta^2>0$, the particle content now consists of
a massive neutral scalar particle and two massive gauge particles. In 
this case vortex solutions are also present. 
Gauss law implies that any vortex solution must have electric  
charge $Q$ (and vice versa): 
\be
\int d^2x\,J^0 =Q=\kappa \Phi\,,
\ee
where $\Phi=\int d^2xF_{12}$ is the flux and $J^{\mu}$ 
is the conserved electro-magnetic current. 
Both vortices and charged particles are anyons.
The first solutions were proposed by Paul and Khare,
\cite{PKh}, a radially symmetric charge one vortex is of the form
\be\label{PK}
\phi=f(r)e^{i\theta}\;,\;\;\;\;
{\bf A}(r)=-\frac{A(r)}{r}\hat{e}_{\theta}\;,\;\;\;
A_0(r,\theta)=A_0(r)\,.
\ee
$f(r)$ and $A(r)$ satisfy boundary conditions required by finite energy
and $A_0(r)$ satisfies $A_0(0)=0$ and $A(\infty)=0$. 
$f(r)$, $A(r)$ and $A_0(r)$ can be solved for numerically. 

The Wilson loop operator is defined by
\be
W(C)=exp(iq\oint_CA_idx^i)\, ,
\ee
where $C$ is a closed oriented loop in Euclidean three-space,
$x^i(s)$, and $q$ a real number. The Wilson loop measures the 
potential between external electric charges of charge $q$. 
The 't Hooft loop operator \cite{tHooft}, $B(C)$,
is the dual object to the Wilson loop operator. 
Its expectation value measures the potential between 
external flux tube solitons, and especially the vortices. 
Its action 
on a field eigenstate $|A_i(x),\,\phi(x)>$ transforms this state by 
a singular
gauge transformation into 
$|A_i^{\Omega_C}(x),\,\phi^{\Omega_C}(x)>$.
The gauge transform  $\Omega_C(x)$ with
\be\label{omega}
\Omega_C(x)=exp(i\omega_C(x))\,,
\ee
it is defined by the following.  
If another closed curve $C'$ links with $C$ $n$ times in a certain 
direction and $C'$ corresponds to $x^i(\theta)$, $0\leq\theta\leq 2\pi$, 
then 
$\omega_C(x(2\pi))=\omega_C(x(0))+2\pi n$.
The singular gauge transform creates a flux tube along the curve $C$,
as measured by $\int_{C'}A_i^{\Omega_C}(x)dx^i$. 
Thus $B(C)$ creates an infinitely thin bare flux tube along the curve $C$. 
To calculate its expectation value it is necessary sum over all field 
configurations which include this infinitely thin 
flux tube, and each configuration of action $A$
is weighted with $e^{-A}$. 
Is it not possible to vary the flux of the vortex/anti-vortex pair 
(except in integer multiples) that is created by the 't Hooft loop, unlike 
the Wilson loop where the charge of the external particle $q$ can be 
arbitrary. The flux of the bare vortices is always 
topologically quantized in integer multiples of $2\pi/e$. 

Perturbatively the expectation values of both $W(C)$ and $B(C)$ have 
perimeter law behaviour for large curves $C$. We take $C$ to be 
rectangular of lengths $t$, and $r$, 
with $t\gg r$ then their expectation values have the 
following behaviour, 
\be
<W(C)>\approx e^{-V(r)t}\;,\quad<B(C)>\approx e^{-V'(r)t}\quad .
\ee 
$V(r)$ and $V'(r)$ measure respectively 
the potential between external charges and fluxes. 
The zero order approximation for $V(r)$ can be shown to reproduce the 
classical electrostatic potential in two dimensions, which for large $r$ 
behaves as $ln\,r$
for a massless gauge theory and $e^{-m_Ar}/(m_Ar)^{1/2}$ for a massive gauge 
theory where $m_A$ is the gauge mass. This remains a good approximation 
for weak coupling, defined by  $e^2\ll M$, 
where $M$ is the smallest mass in the theory.
For the 't Hooft loop, $V'(r)\approx E$ where $E$ is the energy of the created 
object, in this case $E\approx 2m$ where
$m$ is the mass of the vortex. This implies the 't Hooft loop has
perimeter law behaviour perturbatively 
recovering the fact that classically there are 
no long range inter-vortex forces. 

To go beyond the perturbative approximation for the 
Wilson and 't Hooft loops, 
first we ask which configurations are likely to give an important 
contribution to the path integral. The obvious guess to this are the 
vortex loop configurations. As mentioned in the introduction, 
with $\kappa=0$ 
these loops are just Nielson-Olesen vortex loops, but including 
the CS term the form of the vortex solutions in Euclidean 
space is somewhat unclear. 
The CS term becomes imaginary in Euclidean 
space while the remaining action is real.
Classical solutions will typically involve complex valued fields,
the theory however is defined by the functional 
integral over real-valued fields.
Our choice of path integral excludes any solutions 
to the field equations and thus any true instanton solutions. 
To exploit complex critical field 
configurations in the path integral requires analytic continuation of  
the contour in the infinite dimensional space of field configurations.
Here we avoid this approach and restrict the path integral to real fields. 
If we take $\kappa$ to be small then it is 
reasonable to assume that the same 
configurations that are important for the path integral in the absence 
of $\kappa$ will again be important here. This amounts to evaluating the 
action for $\kappa\neq 0$ with the vortex loops which were the instanton 
configurations solutions when $\kappa=0$. 

A similar approach was taken in \cite{AHPS}, concerning
2+1 dimensional compact QED with a CS term. 
The authors assume weak coupling 
and treat the CS term as a perturbation.
This model without the CS term 
was first considered by Polyakov,
\cite{Polyakov}, who showed that the theory is linearly confining because of
instantons. In this case the instantons which are monopoles,
change the logarithmic
perturbative potential between electric charges into a linear confining 
potential. Using the assumption that one may treat 
the CS term as a perturbation the authors of \cite{AHPS} showed that linear 
confinement no longer holds. 
In the instanton calculation, when the CS term is 
evaluated for the monopole the result turns out to depend on the 
$U(1)$ phase of the monopole. When integrating over all
possible monopole configurations in the path integral, 
one integrates over the 
$U(1)$ phases of the monopoles and these 
interfere destructively, cancelling their contribution to the path integral. 
In addition, since the photon is massive the 
perturbative contribution to the potential
between external charges falls off exponentially with their separation. 
It is thus argued in that the 
confinement of electric charges does not hold 
when the CS term is included.
In \cite{Complex}, the above approach of inserting  
instanton solutions from the $\kappa=0$ case into the action for
$\kappa\neq 0$ is questioned. The authors of \cite{Complex} 
find complex monopole solutions with finite action and argue that
that these are the relevant configurations which dominate the path integral,
and are not cancelled by integrating over their phase variables.
We feel that the treatment of the complex solutions 
in the path integral is as yet unclear since a prescription is required on 
how to analytically continue the field configuration space, and 
the question of gauge fixing remains problematic. Nevertheless this 
approach clearly deserves more attention.

Complex vortex solutions are easily seen to 
exist in the present case, although the above 
controversy notwithstanding, we will not use them.  
We expect the situation to be less complicated 
than that of monopoles in the compact $U(1)$ case . This is 
because the vortex field strength vanishes exponentially quickly 
outside the vortex core so we do not have troublesome boundary 
terms which could result in gauge dependence.
We take the same ansatz for the vortex in Euclidean space as for that in 
Minkowski space considered by Paul and Khare, (\ref{PK}).
With the obvious replacement $A_0\rightarrow A_0'=iA_0$ the ansatz
satisfies the Euclidean equations. This is just a consequence of the fact 
that the vortex is time independent. A straight vortex 
line results by assuming independence on $x_3$. The solution has 
finite real action per unit length. However the action per unit length 
is less than the energy of the Minkowski vortex because $A'_0$ is 
imaginary. The action density is given by,
\bea
&&\left\{\frac{1}{2}(\frac{df}{dr})^2+\frac{(1+eA)^2}{2r^2}f^2
+\frac{1}{2r^2}(\frac{dA}{dr})^2+\frac{\lambda}{4}(f^2-\eta^2)^2\right\}\\
&-&\left\{\frac{\kappa}{2}(A\frac{dA_0}{dr}-A_0\frac{dA}{dr})
+\frac{1}{2}(\frac{dA_0}{dr})^2+
\frac{e^2}{2}A_0^2f^2\right\}\,,\nonumber
\eea 
where $A,\,A_0,\,f$ are the same as in (\ref{PK}).   
This differs from the energy of the Paul and 
Khare vortex in that the terms in the second line
are negative, and that the CS term contributes  
to the Euclidean action, but not to the
energy of the Minkowski vortex. The action per unit length 
is less than the energy of the Paul and Khare vortex by terms of 
$O(\kappa^2)$. For large $\kappa$ the 
action per unit length becomes negative.
We expect here that the approximation of 
treating the CS term as a perturbation
fails badly, but for small $\kappa$ we expect it to work well.  

\section{Evaluation of the Wilson and 't Hooft loops }
\news
\ \indent

We now outline the calculation of 
the vortex loop contribution to the expectation values
of the Wilson and 't Hooft loops for $\kappa\neq 0$.
The path integral is split into a integral over all vortex 
loops and the perturbative fluctuations from each 
loop. It is not possible to evaluate the 
exact contribution of the perturbative fluctuations, they give  
a determinant factor, which we presume is not important to our analysis.
The contribution of the vortex loops to the path integral can be evaluated by
approximating each loop as closed string whose 
location is given by the position of the zeros of 
the Higgs field along the vortex loop. 
The dominant contribution to the action of a vortex loop
is its length times its
mass per unit length. The loop may 
have an arbitrary shape, however if the loop
twists in an unruly fashion or intersects itself there will be additional 
potential energy. This suggests in addition to the basic term of the mass, 
terms involving the curvature of the loop and a term which describes the 
potential energy of an intersecting loop could be taken into consideration.  
It is in principle possible to determine the 
detailed form of the interaction terms
by referring back to the field theory model and examining 
how the equations for the Nielson-Olesen vortex change if the shape 
of the straight vortex is deformed \cite{Gr}. It is also necessary to
include terms 
describing the interactions between different loops. Since for a vortex loop
the fields approach their vacuum expectation values exponentially 
quickly in the transverse directions, the only 
long range forces are those mediated by the CS term.

It is well known that the functional integral over a gas of loops 
corresponds to a field theoretic path integral, 
\cite{BarSam}, \cite{ST}, \cite{Sam}. This equivalence
is demonstrated and well studied in the above papers, we will 
not review it here, but just adapt it for our present needs, 
where, especially, $\kappa\neq 0$. For the 
contribution of the vortex loops to the
partition function $e^{-S_E}$,
one gets the following field theoretic path integral,
\be\label{35}
{\cal L}={\cal N}\int{\cal D}\chi\,{\cal D}\chi^*
\,exp\{-\int[\partial_{i}\chi\partial_{i}\chi^*
+m_0^2\,\chi\chi^*+\V]d^3x\}\,,
\ee
with ${\cal N}$ an normalisation factor and 
$\chi$ is the charged vortex field. This corresponds to 
an effective Lagrangian for the vortex field 
$\chi$ which can be treated semi-classically.
The mass term for the vortex field, 
$m^2_0$, has two contributing factors, one from the mass of the 
Nielson-Olesen vortex and an opposing factor due to the entropy factor 
of all possible vortex loops. Working on a fixed cubic lattice, 
for example \cite{ST}, the combination $m-ln\,\mu$ naturally 
emerges as being proportional to the mass $m_0^2$, 
where $m$ is the vortex mass per unit length 
and, $\mu\approx 5$, 
comes from the number of available directions the vortex path can take.
A phase transition will occur at $m=ln\,\mu$, where loops 
of all lengths become important. A similar transition will occur 
here for some value of the parameters. Once the vortex mass becomes 
small enough loops of all lengths are important and the vacuum 
becomes full of vortex/anti-vortex pairs. This is the 
spaghetti vacuum scenario discussed in \cite{BKM}.  

The exact form of the potential $\V$ depends 
on the vortex interactions. For example, repulsive delta functions 
interactions between the vortex loops which prohibit the loops from
intersecting, yields a $\lambda(\chi\chi^*)^2$ potential.
To take into account all the correct interaction terms it is 
necessary to consider the vortex equations.
Vector forces between different part 
of the vortex loop, and different loops, 
necessitates the introduction of auxiliary vector fields which couple 
in a gauge invariant manner to the vortex scalar field, $\chi$. The form 
of the field strength depends on the interaction, for example QED$_3$
arises from Biot-Savart forces between different elements of the vortex 
loop. 
The vector fields, and other possible scalar fields, are then
integrated out to yield the vortex potential $\V$
which is in general nonlocal.

For non-zero $\kappa$ most of the above discussion goes through unaltered.
We saw previously that the vortex solutions still exist when $\kappa\neq 0$ 
in Minkowski space, in Euclidean space any solutions are necessarily complex.
We insert the Euclidean vortex configurations found at $\kappa=0$  
into the action when $\kappa\neq 0$, which should be a reasonable
approximation for $\kappa$ small.  
We can assume that all vortex 
loops arise from vortex/anti-vortex pairs with flux $2\pi/e$, since 
loops containing a multiple of this flux are exponentially suppressed. 
It is not difficult to see that evaluating the CS term for vortex 
loop configurations measures the linking numbers between different 
loops. For two non-intersecting vortex loops evaluating 
$i\kappa/4\int d^3x\epsilon_{ijk}A_{i}F_{jk}$, gives to leading order 
$i\pi^2\kappa n/e^2$ where $n$ is  their linking number. So the 
effect of inserting Nielson-Olesen vortex loops into the complex action is to
multiply their action by a phase which measures how the vortex loops are 
inter-linked. A local expression for $n(C,C')$  is given by
\be\label{pop}
n(C,C')=-\frac{1}{4\pi}\int_Cd{\bf x}\cdot\int_{C'}\frac{({\bf x}-{\bf y})
\times d{\bf y}}{|{\bf x-y}|^3}\,.
\ee
The linking number is thus manifested as a vector interaction 
between the vortex loops.  Combined with the previous vector interactions
from above, this will yield a vector field $G_i$ coupling in a gauge 
invariant manner to the vortex field $\chi$. The field strength will now 
have a complex part, due to the fact that the linking number interaction
is imaginary. For example, combining the linking number 
interaction with the Biot-Savart interaction yields a complicated nonlocal 
field strength whose lowest order terms in $\kappa$ are,
\be
\frac{1}{2}\left\{{\bf H}^2+i\frac{\kappa\pi}{4(ee')^2}{\bf H}
\cdot\nabla\times{\bf H}\right\}+\dots\quad,
\ee
where ${\bf H}=\nabla\times{\bf G}$, and $e'$ is a parameter which 
measures the strength of the Biot-Savart interaction.
Again the correct procedure is to integrate out the vector field 
$G_i$, yielding (\ref{35}). It is important that $\V$ depends only 
on $\chi\chi^*$, and not on $\chi$ or $\chi^*$ separately.
This is due to the global symmetry $\chi\rightarrow e^{i\theta}\chi$ which 
corresponds to conserved vortex number. 
The results below are not 
sensitive to the exact form of $\V$. The mechanism of how the spaghetti 
vacuum occurs is unchanged relative to the $\kappa=0$ case.
When the parameter $m^2_0$ in (\ref{35}) becomes negative the vortex 
field will acquire a nonzero vev. The vortex loops are stabilised 
by terms of the form $\lambda(\chi\chi^*)^2$ in $\V$ arising from the 
repulsive forces. We denote the critical value of $\eta^2$
($\approx m$, the vortex mass), where the phase transition occurs  
as $\eta^2_{crit}$.

To evaluate $<W(C)>=<exp(iq\oint_CA_idx_i)>$, 
(here $A_i$ is the gauge field of the original theory), note that 
$\oint_CA_idx_i$ measures $2\pi/e$ times the linking number of 
a vortex loop configuration, $n(C)$, with the curve $C$ and has 
the integral expression,
\be
n(C)=\oint_{C}dx_iP_i(x)\,,
\ee   
where 
\be\label{wil}
P_i(x)=-\frac{1}{4\pi}\sum_{C'}
\int_{C'}\frac{\epsilon_{ijk}(x-y)^j
dy^k}{|x-y|^3}\,.
\ee
The sum over $C'$ represents the sum over the vortex loops in a 
particular field configuration. We emphasise that this linking number 
does not come from the CS term.
Thus to calculate the expectation 
value we do the same sum as before but now include the linking number 
factor. This essentially gives the complex scalar field theory in the 
external gauge field $P_i$
which the Wilson loop generates, \cite{Sam}, \cite{ST}. We get
\be\label{312}
{\cal N}\int{\cal D}\chi\,{\cal D}\chi^* 
exp\{-\int[D_{i}\chi D_{i}\chi^*
+m_0^2\,\chi\chi^*+{\cal V}(\chi\chi^*)]d^3x\}\,,
\ee
where $D_{i}=\partial_{\mu}-2\pi iq/eP_i$. 

Note that there is no field strength 
for $P_{i}$, it is just an external field. (\ref{312}) is to be evaluated  
semi-classically. When $m_0^2>0$, $<\chi>=0$ is expected for 
the vacuum and the topological symmetry is unbroken. $\chi=0$ is a 
solution to the equations of motion even for non-zero $P_{i}$ and the 
Wilson loop is of order 1. Vortex loops do not contribute appreciably 
to the Wilson loop. When $m_0^2<0$ however, 
$\chi$ will have a non-zero vacuum expectation
value, which is called topological symmetry breaking  $<\chi>=\chi_0$. 
Inserting this as a trial solution into the path integral gives
\be\label{313}
<W(C)>=<exp[iq\oint_C A\cdot dx]>\approx exp[-(\frac{2\pi q}
{e})^2\chi_0^2\;ln(\frac{r}{r_0})]\,, 
\ee
where again $C$ is rectangular with lengths $r$ and $t$, $t\gg r$, 
with $r$ representing the separation of the electric charges, and 
$r_0$ is a cut off of the order of the vortex width.
For $q>e$ non-constant screening 
solutions provide a better minimisation 
of the action \cite{Sam}. The upshot is that $q/e$ in 
(\ref{313}) is replaced by $\Delta q=q/e-n$
where $n$ is the nearest integer to $q/e$.
The Wilson loop has logarithmic dependence on the charge separation 
and is periodic in the external charge $q$ with period $e$, which
evinces the fact that when the spaghetti vacuum forms we are back in 
the region of unbroken $U(1)$ or symmetry unbroken 
phase: there exist elementary
charged particles of charge $e$ which can screen external particles of 
charge $e$ or an integer multiple thereof. 
This is not a conclusion that one would 
have expected. With a massive photon one would not expect any form of
confinement, as is the case for compact $U(1)$ with a CS term
\cite{AHPS}, \cite{KK}. The essential difference between the two cases is 
that is that when evaluating the CS term for the monopole, it explicitly 
depends on the monopole phase, and integrating over these phase variables 
the monopole contribution cancels. In the present case, the effect of
evaluating the CS term for a configuration of vortex loops gives a phase 
proportional to the linking numbers of the different loops.
Given a gas of such vortex loops there is no reason to expect these 
phases cancel. This is manifested by the fact 
that $\V$ depends only on $\chi^*\chi$. 

For large values of $\eta^2$ the vortex loops only
give a small contribution to the path integral due to their large mass. As
$\eta^2$ is decreased vortex loops become more important until there is a 
phase transition at $\eta^2_{crit}$ where loops of all length become 
important and the $U(1)$ symmetry is restored. For $\eta^2>\eta^2_{crit}$
the Wilson loop has perimeter law behaviour. The perturbative contributions 
give a potential that falls off exponentially quickly between external 
charges. The nonperturbative vortex loop configurations may alter this 
potential, arising from short vortex loops 
which link or intersect the Wilson 
loop. But their exact contribution to the potential is undetermined 
from this approach since when the vortex loop intersects the Wilson 
loop there is no simple answer for the contribution 
to the expectation value. 

For $\eta^2<0$ the Nielson-Olesen vortex loop 
configurations do not exist. The question is whether flux 
tube configurations can be important when $\eta^2<0$. 
We believe that vortex configurations continue to be relevant to the 
vacuum structure throughout the unbroken symmety phase even though they 
are absent classically. In \cite{KR} for the Abelian-Higgs model 
it is shown that the vortex operator 
possesses a non-zero expectation value throughout the Coulomb phase.
Numerical evidence presented in \cite{Loop} indicates that the vortex 
condensate persists throughout the Coulomb regime. This suggests that 
the confining behaviour will persist throughout the unbroken 
symmetry phase. 

Next we turn to the evaluation of the 't Hooft loop defined along a curve $C$ 
as before.  Now the complex part of 
the Euclidean action plays a crucial role. To calculate the 't Hooft 
loop expectation value we must sum over all field configurations which 
contain an infinitely thin flux tube along $C$. This is equivalent to 
calculating the partition function, $e^{-S_E}$, where the functional 
integral is over all gauge configurations $A_i'$, such that,
$A_i'=A_i+\partial_i\Omega_C$ with $\Omega_C$ as in (\ref{omega}) and 
$A_i$ is non-singular.
In the absence of the CS term, because the vortex 
loops have no long range interactions, the 't Hooft loop will have 
perimeter law dependence.
Inclusion of the CS term however implies that for each 
vortex loop $C'$ that links $n(C,C')$ times with $C$ we get a term, 
\be
exp\,\{i\kappa(\frac{2\pi}{e})^2n(C,C')\}\,,
\ee
where $2\pi/e$ is the flux carried by 
the 't Hooft loop $C$. This in turn means that 
the 't Hooft loop calculation is 
identical to the Wilson loop calculation with the replacement 
$q\rightarrow 2\pi\kappa/e$, which due to the Gauss law, is nothing else than 
the electric charge of a vortex that has flux $2\pi/e$. 
Hence in the region $\eta^2<\eta^2_{crit}$, vortex 
loops give a logarithmic contribution to the 't Hooft loop
expectation value and one 
expects logarithmic confinement of external 
vortices. Such confinement of vortices has 
already been seen to occur in a related model \cite{DST1}.
The previous remarks in the last paragraph about the validity of this 
calculation when $\eta^2<0$ also apply here.
If the external vortices have electric 
charge equal to a multiple of $e$ then the logarithmic 
potential disappears, however we understand this in terms 
of screening by the elementary charged scalar particles of 
the theory.  Thus it is actually the electric charge of the vortices
that is behind the confinement.  It is again clear that we are in 
the region of unbroken symmetry
here, since for long range forces to exist among the external 
vortices, the dynamical vortices must have disappeared from the theory. 

Electrically charged particles of charge $q\,\mbox{mod}\,e\neq 0$ are 
logarithmically confined in the topological symmetry breaking phase 
(or spagetti vacuum) as discovered by Samuel \cite{Sam}. Our new 
result is that this remains true with the addition of the 
CS term, at least for small CS coefficient. 
Additionally, this includes the logarithmic 
confinement of vortices 
since they also carry electric charge due to the Gauss law.
This result is surprising in view of the fact that all gauge fields
are massive and interactions short ranged and demonstrates the
importance of vortex configurations for the phase structure of the
theory. As a note of caution, we stress that the methods 
employed here are far from rigorous. The procedure of obtaining an 
effective vortex Lagrangian is more problematic 
than when considering, for example,
monopoles or other point-like instanton configurations, 
with a well defined set of 
zero modes. Obtaining the vortex action here, by summing over 
all string configurations has intrinsic 
ambiguities. Finally the question of complex field configurations
and their relevance in the path integral needs to be examined in more detail.  
\\[10pt]
\noindent{\bf Acknowledgement}
We thank NSERC of Canada and FCAR of Qu\'{e}bec for financial support. 
We thank R. B. Mackenzie and V. Rubakov for useful discussions. 
\\[5pt]

\end{document}